# Complete phase diagram of Sr$_{1-x}$La$_x$FeO$_3$ with versatile magnetic and charge ordering

M. Onose[1,2], H. Takahashi[1,2,†], H. Sagayama[4], Y. Yamasaki[5], and S. Ishiwata[1,2,3]

[1] *Department of Applied Physics, University of Tokyo, Bunkyo-ku, Tokyo 113-8656, Japan*

[2] *Division of Materials Physics, Graduate School of Engineering Science, Osaka University, Toyonaka, Osaka 560-8531, Japan*

[3] *Center for Spintronics Research Network (CSRN), Graduate School of Engineering Science, Osaka University, Toyonaka, Osaka 560-8531, Japan*

[4] *Institute of Materials Structure Science (IMSS), High Energy Accelerator Research Organization (KEK), Tsukuba, Ibaraki 305-0801, Japan*

[5] *National Institute for Materials Science (NIMS), Tsukuba, Ibaraki 305-0047, Japan*

## Abstract

A detailed electronic phase diagram of perovskite-type oxides Sr$_{1-x}$La$_x$FeO$_3$ ($0 \leq x \leq 0.5$) was established by synchrotron X-ray diffraction, magnetization, and transport measurements for polycrystalline samples synthesized by a high-pressure technique. Among three kinds of helimagnetic phases in SrFeO$_3$ at zero field, two of them showing multiple-$q$ helimagnetic spin textures tend to rapidly disappear in cubic symmetry upon the La substitution with $x$ less than 0.1, which accompanies the loss of metallic nature. On the other hand, the third helimagnetic phase apparently remains robustly in Sr$_{1-x}$La$_x$FeO$_3$ with $x$ higher than 0.1, followed by merging to the spin/charge ordered phase with $x \sim 1/3$. We propose an important role of itinerant ligand holes on the emergence of multiple-$q$ states and a possible link between the third (putative single-$q$) helimagnetic phase in SrFeO$_3$ and the spin/charge ordered phase in Sr$_{2/3}$La$_{1/3}$FeO$_3$.

## Introduction

Spin/charge ordering phenomena in 3$d$ transition-metal oxides are the manifestation of the emergent interplay between spin and charge degrees of freedom in strongly correlated electron systems.[1–6] Perovskite-type ferrites with unusually high valence Fe ions are such attractive oxides showing a rich variety of unique spin/charge ordering.[7,8] For instance, as for perovskites $A$FeO$_3$ ($A$ = Ca, Sr) with Fe$^{4+}$ ions, CaFeO$_3$ shows incommensurate helimagnetic ordering and charge ordering due to charge disproportionation of Fe$^{4+}$ into Fe$^{3+}$ and Fe$^{5+}$,[9] while SrFeO$_3$ has been revealed to host various helimagnetic structures with being free from charge ordering instability.[10] On the other hand, for perovskites with Fe$^{3.67+}$ ions, Sr$_{2/3}$La$_{1/3}$FeO$_3$ with rhombohedral distortion has been reported to show a commensurate spin/charge ordering; three-fold charge ordering as -Fe$^{3+}$-Fe$^{3+}$-Fe$^{5+}$- and six-fold spin arrangement as ↑↑↑↓↓↓, both propagating along the [111] direction in the pseudo-cubic unit cell.[11–13]

Recently, the spin/charge ordering in Sr$_{2/3}$La$_{1/3}$FeO$_3$ has been revisited and found to be potentially different from the one originally proposed by Battle *et al.*[11] As for the charge ordering, Blasco *et al.* have proposed the charge distribution of Fe$^{3.67+\delta}$-Fe$^{3.67+\delta}$-Fe$^{3.67-2\delta}$ rather than Fe$^{3.67-\delta}$-Fe$^{3.67-\delta}$-Fe$^{3.67+2\delta}$ (-Fe$^{3+}$-Fe$^{3+}$-Fe$^{5+}$-) on the basis of X-ray absorption spectroscopic analyses.[14] They also suggested that an order-disorder transition takes place therein, instead of the significant charge disproportionation. Furthermore, a neutron diffraction experiment has suggested that the helimagnetic spin structure with $q$ = (1/6, 1/6, 1/6) in the pseudo-cubic cell can be eligible as well as the collinear spin structure.[15] It should be noted here that Ca$_2$La$_{1/3}$FeO$_3$ with A-site ordered perovskite structure has recently been found to show two types of spin/charge ordering as a function of temperature.[16,17] Whereas the spin/charge ordering at lower temperatures is the same as that of Sr$_{2/3}$La$_{1/3}$FeO$_3$, the other at higher temperatures has charge ordering along [010] and helical spin structure with a propagation vector $q$ = (1/3, 1/3, 1/3) on the pseudo-cubic unit cell. These recent progress on $A_{2/3}$La$_{1/3}$FeO$_3$ implies that the ground states of perovskite-type ferrites with Fe$^{3+/4+}$ ions remain to be established.

In this work, we systematically studied structural, magnetic, and electronic properties of Sr$_{1-x}$La$_x$FeO$_3$ with $x$ from 0 through 0.5 to see how the helimagnetic phases at $x$ = 0 change to the spin/charge ordered phase at $x$ = 1/3, and eventually to the antiferromagnetic Mott insulating phase in the La-rich region. A cubic perovskite SrFeO$_3$ has been reported to host five different helimagnetic phases, termed phase I through V.[10] Phases I through III are stable at zero field, whereas phases IV and V appear only in the presence of the magnetic field. Recent neutron diffraction experiments have revealed that phases I and II can be characterized by the anisotropic double-$q$ and isotropic quadruple-$q$ spin spirals propagating along the <111> equivalents, respectively, the latter of which yield three-dimensional topological spin texture with hedgehog and anti-hedgehog singularities.[18] Phase III has also been found to be helimagnetic with the spin propagation along the <111> direction, but it remains elusive how the spins are twisting and whether the helical spin modulation vector is single or multiple. In this study, we discuss the La-doping effect on these helimagnetic phases and the relation between phase III and the commensurate spin/charge ordered phase in Sr$_{2/3}$La$_{1/3}$FeO$_3$.

## Experimental Methods

Polycrystalline samples of Sr$_{1-x}$La$_x$FeO$_3$ were prepared by the following two-step method. First, polycrystalline samples of the oxygen deficient perovskites Sr$_{1-x}$La$_x$FeO$_{3-\delta}$ were synthesized by a conventional solid-state reaction. The starting materials (SrCO$_3$, La$_2$O$_3$, and Fe$_2$O$_3$) were stoichiometrically mixed and sintered at 900-1250 °C in air for 12-24 h. This process was repeated 2-3 times. The obtained powder was pelletized and sintered again at 1300 °C for 24 h in a flow of Ar gas. Second, the sintered sample was sealed in a gold capsule with an oxidizing agent ($NaClO_3$), and annealed at 500 °C and 8 GPa for 1 h using a cubic-anvil-type high-pressure apparatus.

Synchrotron powder X-ray diffraction (XRD) was performed at beamline 8A, Photon Factory, KEK, Japan. The wavelengths of synchrotron X-ray in the range from 0.68975 Å to 0.68987 Å were calibrated for each compound by a standard sample of $CeO_2$. Crystal structures were refined by Rietveld analyses using RIETAN-FP.[19] Bond valence sum (BVS) for the Fe ion was calculated by BVS = $\Sigma_i \exp((r_0 - r_i)/0.37)$, where $r_i$ represents the Fe-O bond length. $r_0$ for each composition was obtained by the linear interpolation between 1.759 ($Fe^{3+}$, $x = 1$) and 1.78 ($Fe^{4+}$, $x = 0$).[20] Magnetic properties and resistivity were measured with MPMS and PPMS (Quantum Design), respectively.

## Results and Discussion

Figures 1(a) and 1(b) show the synchrotron powder XRD patterns of $Sr_{1-x}La_xFeO_3$ with selected compositions at room temperature and their magnified view around $2\theta \sim 17°$, respectively. The results of Rietveld refinement for compositions with $x = 0.25$ and 0.45 are shown in Figs. 1(c) and 1(d), and the refined structural parameters are summarized in Table 1. For $0 \leqq x \leqq 0.25$, all reflections can be indexed well with a simple cubic perovskite-type structure ($Pm\text{-}3m$). For $0.33 \leqq x \leqq 0.5$, the reflection at $2\theta \sim 17°$, which is absent in the cubic phase, can be indexed as 113 in the rhombohedral structure ($R\text{-}3c$), as previously reported.[14,21] Figs. 2(a)-(d) show the $x$ dependence of lattice constant, the Fe-O bond length, the Fe-O-Fe bond angle, and BVS of Fe ions, respectively. For the rhombohedral phase with $0.33 \leqq x \leqq 0.5$, the lattice constant is converted to the pseudo-cubic values. Upon the La doping for $SrFeO_3$, the lattice constant increases linearly following the Vegard's law, indicating that the oxygen deficiency of the samples is negligible. With increasing $x$, the Fe-O bond length increases monotonically, reflecting the decrease in the valence of Fe ions from +4 to +(4-$x$) as oxidized in BVS for Fe ions, while the average ionic radius of A-site ion decreases. Thus, the La doping is expected to reduce the tolerance factor from unity, which manifests itself as the structural phase transition from cubic ($0 \leqq x \leqq 0.25$) to rhombohedral ($0.33 \leqq x \leqq 0.5$) and the decrease in the Fe-O-Fe bond angle down to 168° at $x = 0.5$ in the rhombohedral phase. From the interpolation of the deviation of the Fe-O-Fe bond angle from 180°, the critical composition for the structural transition can be estimated as $x \sim 0.3$.

Figures 3(a)-(d) show the temperature dependence of magnetic susceptibility and resistivity. As shown in Figs. 3(a)-(c), a cusp indicating an onset of antiferromagnetic order can be found for all compositions in the temperature-increasing process after zero-field cooling (ZFC). Upon increasing $x$, the temperature dependence of resistivity shown in Fig. 3(d) exhibits a systematic change from metallic to insulating, whereas the absolute value of resistivity develops non-monotonously around $x = 0.17$, 0.2, 0.25, potentially due to the experimental uncertainty in the resistivity measurements for polycrystalline samples. For $0.33 \leqq x \leqq 0.4$, a kink or a jump in resistivity is observed at the antiferromagnetic ordering temperature, which most likely corresponds to the spin/charge ordering as reported previously.[14] It should be noted that the compounds with $x = 0.17$, 0.2 and 0.25 show a smeared kink in the resistivity upon the magnetic ordering, which will be discussed later. As $x$ increases from 0.33, a large difference between FC (field cooling) and ZFC runs appears below the magnetic transition temperature (see also Fig. 6(a)), which is most likely attributed to weak ferromagnetism induced by the lattice distortion. The enhancement of the difference between FC and ZFC curves with La doping presumably reflects the development of spin canting due to the increase in the rhombohedral distortion as confirmed in Fig. 2(c).

Figures 4(a)-(c) show the magnified view of Figs. 3(a) and 3(d) for $0 \leqq x \leqq 0.1$. For $0 \leqq x \leqq 0.02$ (0.05 $\leqq x \leqq 0.07$), there exists two (one) first order phase transitions below the antiferromagnetic transition temperature around 130K, which is also apparent in the temperature derivative of magnetic susceptibility and resistivity shown in Figs. 4(d)-(f). These transitions correspond to the onset of the three different helimagnetic phases in $SrFeO_3$ at zero field (phase I, II, and III).[10] Whereas the transition temperature for phase III remains almost constant around 130 K, those for phases II and I tend to decrease upon La substitution with $x$ less than 0.1, being accompanied by the loss of metallic character.

Figures 5(a)-(c) show the magnetic field dependence of magnetization (left axis) and magnetization divided by magnetic field (right axis) for selected compositions. Whereas the magnetization curves for $x = 0$, 0.05 are weakly nonlinear (concave downward) with the magnetic field, that for $x = 0.1$ is almost linear, as confirmed by the slope of $M/B$ as a function of $B$ (see also Fig. S1 in the supplemental material). This $x$-dependent change in the magnetization curves may reflect the change in the ground state from phase I to phase III, or the increase in the critical field for the transition from a proper-screw state to a conical state as the period of spin spiral at zero field becomes shorter as will be discussed later. Figs. 5(d)-(i) show the magnetoresistance (MR) defined as MR = $\Delta\rho(B)/\rho(0)$ ($\Delta\rho(B)$= $\rho(B) - \rho(0)$) at 5 K and 100 K. Positive MR is apparent in $x = 0$ at $T = 100$ K, whereas negative MR is observed at other compositions and temperatures (see also Fig. S2 in supplemental material). It has been reported for $SrFeO_3$ that negative MR is observable for all helimagnetic phases except for phase II, where positive MR is found as a characteristic feature for the topological spin texture.[18] In fact, MnGe with a similar topological spin texture shows positive MR, of which origin has been discussed in terms of the large spin fluctuations brought by the field-induced annihilation of hedgehogs and anti-hedgehogs.[22]

Summarizing the magnetic and resistivity data of $Sr_{1-x}La_xFeO_3$, we obtained the electronic phase diagram shown in Fig. 6(c). Depending on the La content $x$, there exist three characteristic regions: the helimagnetic metal phase ($0 \lesssim x \lesssim 0.1$), the spin/charge ordered phase ($0.33 \lesssim x \lesssim 0.4$), and the canted antiferromagnetic insulating phase ($0.45 \lesssim x \lesssim 0.5$). In the helimagnetic metal phase, the significant La-doping effect on phases I and II manifests itself as the subtle sensitivity of multiple-$q$ states to the change in the valence of Fe ions, which will be discussed later. On the other hand, phase III is robust against the La doping and continuously connected to the spin/charge ordered phase at $x = 0.33$ through the transient region of $x = 0.17 \sim 0.25$. Since the magnetic susceptibility for ZFC at $T = 5$ K shows a monotonic decrease with increasing $x$ for $x < 0.33$ and becomes constant for $x \gtrsim 0.33$ (see Fig. 6(a)), the helical spin structure is expected to be collinear somewhere below $x = 0.33$. We note here that this $x$-dependent magnetic transition seemingly coincides with the structural phase boundary of the cubic-rhombohedral transition. To elucidate the relation between the lattice symmetry and the spin structure, it is necessary to perform detailed magnetic studies such as neutron diffraction measurements. Considering the fact that the kinks in magnetic susceptibility and resistivity data for $x = 0.17 \sim 0.25$ are smeared as compared with those at $x = 0.33$ (see Figs. 3(b) and (d)), there remains a possibility that phase III and the spin/charge ordered phase are coexisting in a wide composition range of $x$ down to 0.17 owing to the first-order phase boundary and the chemical disorder. In fact, the previous Mössbauer spectroscopy measurements indicated a slight occurrence of charge ordering even for $x = 0.1$.[13] For $x$ larger than 0.4, the system becomes insulating in all temperature region, which is supposed to be connected with the antiferromagnetic insulating state at $x = 1$.[23]

Next, we discuss the variation of helimagnetic ordering at the low-$x$ region. Although the localized magnetic moment is expected to increase as the metallic state at $x = 0$ with $Fe^{4+}$ ($S = 2$) is replaced by the insulating state at $x = 0.33$ with $Fe^{3.67+}$ ($S = 2.16$), the magnetization at $B = 7$ T decreases systematically with $x$ as shown in Figs. 5(a)-(c) and Figs. S3(a)-(b) in the supplemental materials. This indicates that the La substitution for $SrFeO_3$ tends to suppress the nearest-neighbor ferromagnetic interaction relative to the competing (both nearest-neighbor and further-neighbor) antiferromagnetic interactions. This assumption is reasonable considering that the ligand-hole carriers, which mediate the ferromagnetic nearest-neighbor interaction between $Fe^{4+}$-$Fe^{4+}$, are reduced by the introduction of $Fe^{3+}$ as $x$ increases. Here we assume that the helimagnetic order in $SrFeO_3$ manifests itself as the competition between the ferromagnetic double-exchange interaction and the antiferromagnetic superexchange interaction as proposed for the $Fe^{4+}$-based helimagnet $Sr_3Fe_2O_7$.[24] Considering the fact that the magnetic transition temperature changes smoothly as a function of $x$ from 0 to 0.33 (see Fig. 6(c)), it is reasonable to presume that the incommensurate helical period at $x = 0$ ($q = (0.13, 0.13, 0.13)$)[10] becomes shorter as $x$ increases, eventually coinciding with the commensurate magnetic order at $x = 0.33$. This is an indicative of a single-$q$ spin spiral propagating along [111] for phase III.

Assuming a single-$q$ helimagnetic order for phase III, the effective destabilization of phases I and II by the La substitution in a cubic lattice symmetry can be understood in terms of decrease in itinerant nature of the system. In itinerant magnets with high lattice symmetry, a $pd$ hybridization term[25] or a higher-order perturbation term beyond ordinary RKKY interaction[26] was suggested to play an important role for the emergence of multiple-$q$ helimagnetic order in the strong and weak spin-charge coupling regime, respectively. These two terms are both expected to be enhanced in itinerant magnets with sufficiently high electronic conductivity. In fact, the magnetic ground state in insulating $CaFeO_3$ with monoclinic distortion has been recently confirmed to be single-$q$ spin spiral unlike the case in metallic $SrFeO_3$.[27] In $Sr_{1-x}La_xFeO_3$, the La substitution up to 10 % systematically increases resistivity (see Fig. 6(b)) and change the metallic state to semiconducting state (see Fig. 3(d)) with no influence on the cubic symmetry, meaning that the La substitution causes the reduction of itinerant ligand holes; i.e., the weakening of $pd$ hybridization upon the decrease in the Fe valence state.[23] Therefore, the presence of itinerant ligand holes is essential for the emergence of multiple-$q$ helimagnetic order. This is consistent with the fact that the slight Co doping has no influence on the metallic character and cubic symmetry of $SrFeO_3$ but tends to destabilize phase III rather than phase I and II with multiple-$q$ spin spirals.[28]

## Summary

In summary, we have established the electronic and structural phase diagram of $Sr_{1-x}La_xFeO_3$ ($0 \leq x \leq 0.5$), revealing the effective destabilization of phases I and II with multiple-$q$ helimagnetic orders in $SrFeO_3$ by La doping. The phase diagram and the resistivity data suggest the importance of metallic nature reflecting the presence of itinerant ligand holes for the stabilization of multiple-$q$ helimagnetic order, which is consistent with the theoretical works.[25,26] Furthermore, a signature of the close connection between the spin/charge ordered phase with $x \sim 1/3$ and phase III with presumably single-$q$ spin spiral propagating along [111] was found as the continuous change in the critical temperature and magnetization. Although it is difficult to exclude a possibility that phase III and the spin/charge ordered phase are separated by the first order phase boundary around $x = 0.17 \sim 0.25$, we found a promising clue to identify the spin structure of phase III emerging in a narrow temperature window for $SrFeO_3$.

## Acknowledgements

The authors would like to thank M. Takano and Y. Tokura for valuable comments and experimental supports for cryogenic measurements. This work was partly supported by JSPS, KAKENHI (Grants No. 17H01195, No. 16K17736, No. 17H06137, and No. 19K14652), JST PRESTO Hyper-nano-space design toward Innovative Functionality (Grant No. JPMJPR1412), and Asahi Glass Foundation. The synchrotron powder XRD measurement was performed with the approval of the Photon Factory Program Advisory Committee (Proposal No. 2015S2-007 and No. 2018S2-006).

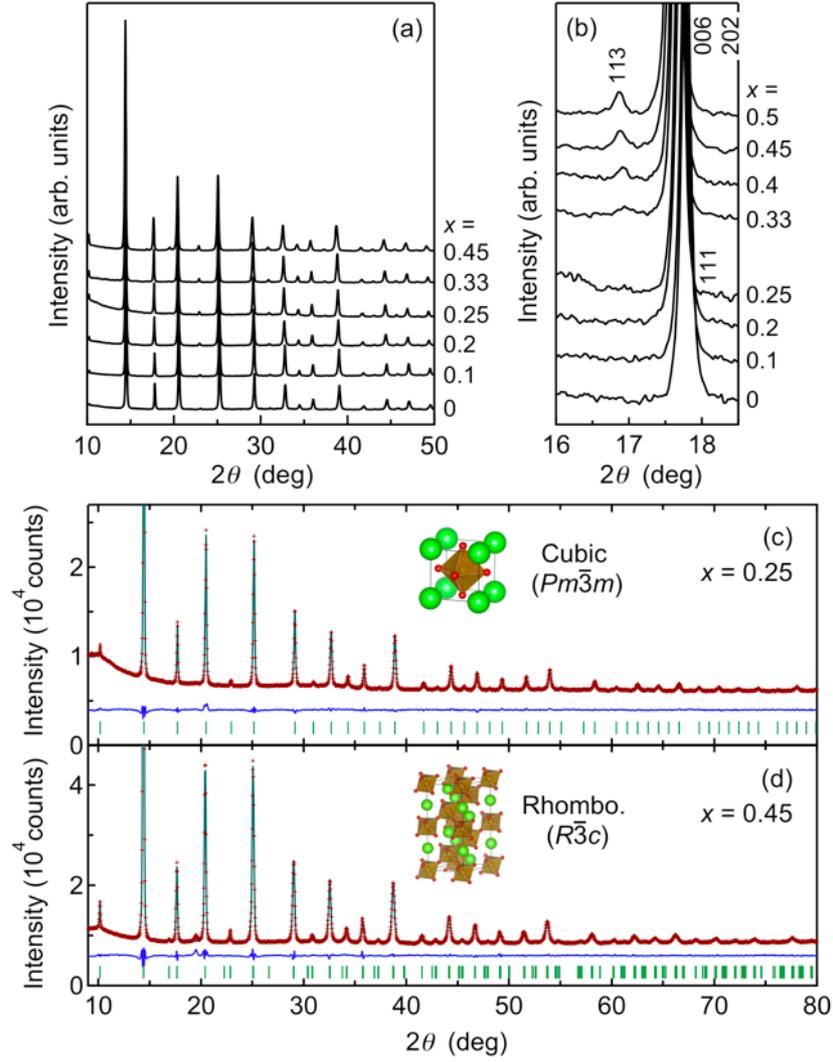

Fig. 1. (a) Synchrotron powder XRD patterns for $Sr_{1-x}La_xFeO_3$ with selected compositions at room temperature and (b) the magnified view around $2\theta$ of 17°. Rietveld refinement plots of synchrotron XRD data for $Sr_{1-x}La_xFeO_3$ with (c) $x = 0.25$ and (d) $x = 0.45$. The red dots, green lines, blue lines, and vertical green tick marks represent observed patterns, calculated patterns, differences between observed and calculated intensities, and Bragg reflection points, respectively. The insets in (c) and (d) show the crystal structure of $x = 0.25$ and $x = 0.45$, respectively.

| $x$ | Space group | Lattice parameters (Å) | Atomic parameters | | | | | | | Reliability factors | |
|---|---|---|---|---|---|---|---|---|---|---|---|
| | | | Atom | Site | Occupancy | $x$ | $y$ | $z$ | $B_{iso}$ (Å²) | $R_{wp}$ | $S$ |
| 0.25 | $Pm$-$3m$ | $a = 3.869(2)$ | Sr | 1$b$ | 0.75 | 1/2 | 1/2 | 1/2 | 0.27(2) | 0.930 | 0.7770 |
| | | | La | 1$b$ | 0.25 | 1/2 | 1/2 | 1/2 | 0.27(2) | | |
| | | | Fe | 1$a$ | 1.0 | 0 | 0 | 0 | 0.13(2) | | |
| | | | O | 3$d$ | 1.0 | 1/2 | 0 | 0 | 0.58(4) | | |
| 0.45 | $R$-$3c$ | $a = 5.497(5)$ $c = 13.437(9)$ | Sr | 6$a$ | 0.55 | 0 | 0 | 1/4 | 0.32(2) | 1.400 | 1.3859 |
| | | | La | 6$a$ | 0.45 | 0 | 0 | 1/4 | 0.32(2) | | |
| | | | Fe | 6$b$ | 1.0 | 0 | 0 | 0 | 0.16(2) | | |
| | | | O | 18$e$ | 1.0 | 0.465(1) | 0 | 1/4 | 0.25(8) | | |

Table. 1. Refined structural parameters for $Sr_{1-x}La_xFeO_3$ with $x = 0.25$ and $x = 0.45$ at room temperature.

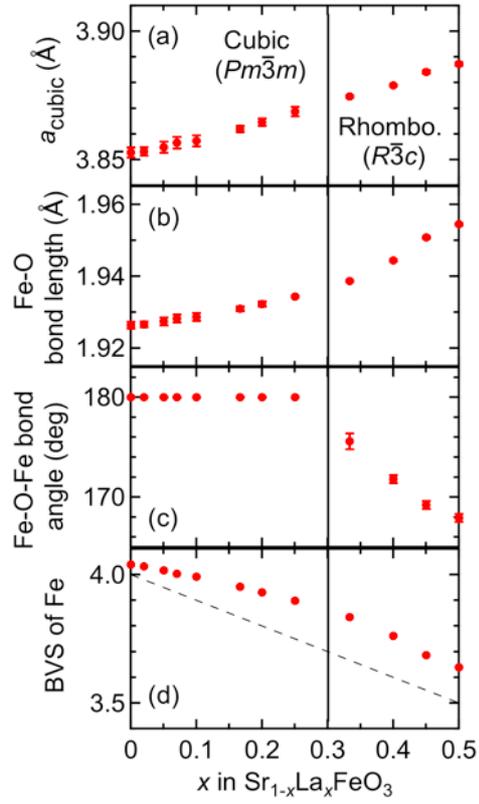

Fig. 2. $x$ dependences of (a) pseudo-cubic lattice constant $a_{cubic}$, (b) Fe-O bond length, (c) Fe-O-Fe bond angle, and (d) bond valence sum (BVS) of Fe ion. $a_{cubic}$ for the rhombohedral compounds with $x \gtrsim 0.33$ was calculated by $a_{cubic} = (V/6)^{1/3}$ ($V$: volume of the rhombohedral unit cell). The broken line in (d) corresponds to the formal valence predicted from the chemical composition.

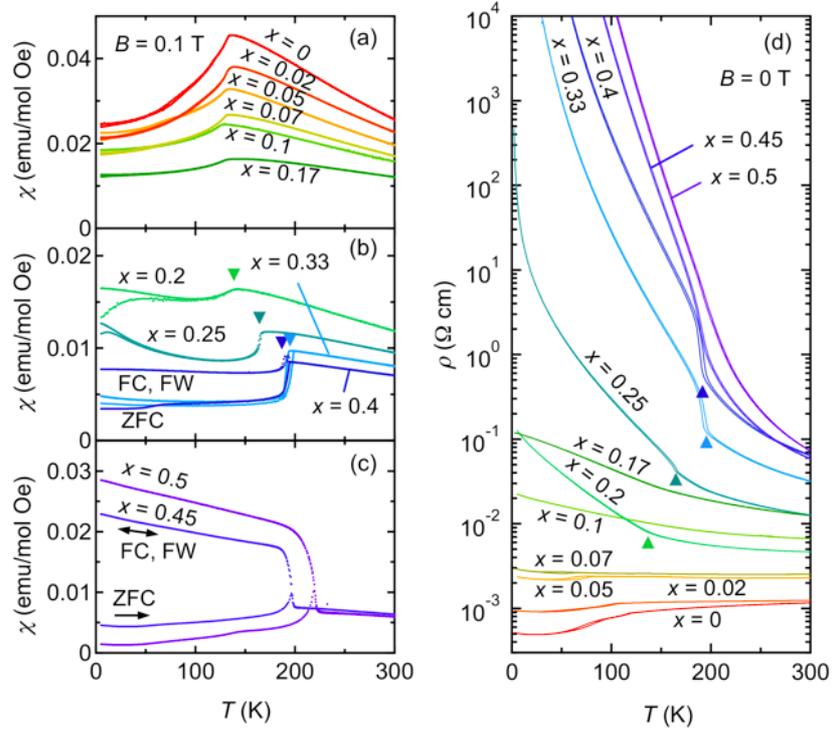

Fig. 3. (a)-(c) Temperature dependence of magnetic susceptibility for $Sr_{1-x}La_xFeO_3$ measured at 0.1 T. The measurements were performed on field cooling (FC), on warming after field cooling at 0.1 T (FW), and on warming after zero-field cooling (ZFC). (d) Temperature dependence of resistivity for $Sr_{1-x}La_xFeO_3$ measured on cooling and heating at 0 T. The closed triangles in (b) and (d) denote the onset of spin/charge ordering.

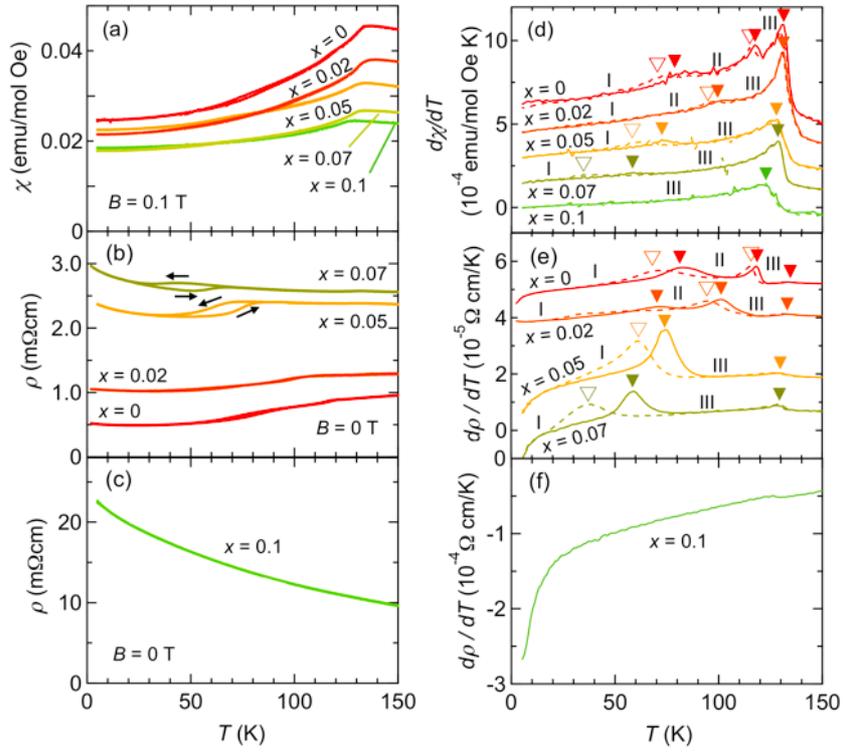

Fig. 4. Temperature dependence of (a) magnetic susceptibility and (b)(c) resistivity for $0 \leqq x \leqq 0.1$ measured on cooling and warming runs. Temperature derivative of (d) magnetic susceptibility and (e)(f) resistivity for $0 \leqq x \leqq 0.1$. The solid and broken lines in (d) and (e) show the data for warming and cooling runs, respectively. The closed and open triangles correspond to the transition temperatures separating the different helimagnetic phases (phases I, II, and III) determined from the data for warming and cooling runs, respectively. In (d) and (e), the data are offset for clarity. For $x = 0.02$, the transition between phase I and II is apparent only in the resistivity data.

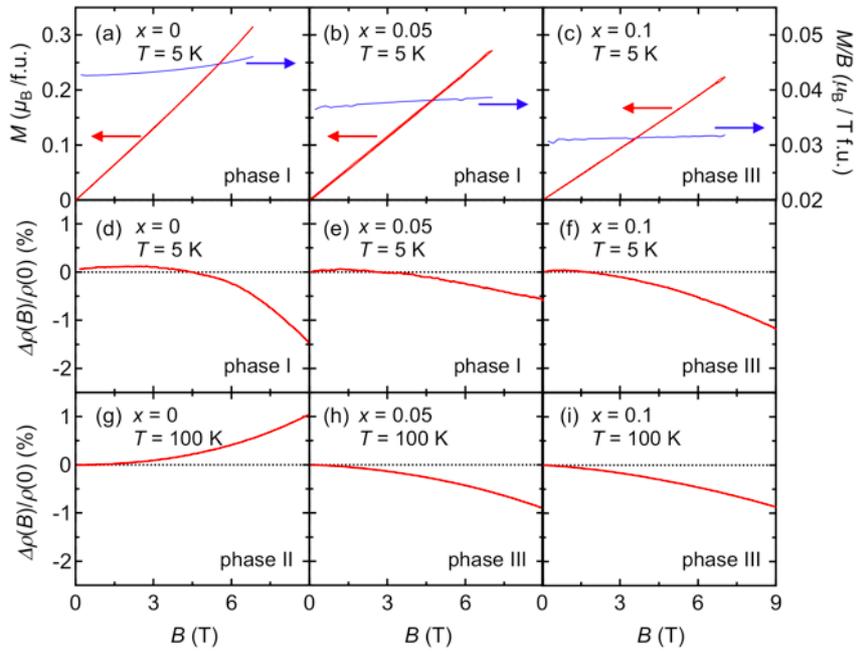

Fig. 5. (a)-(c) Magnetic field dependence of magnetization (left axis) and magnetization divided by magnetic field (right axis) at 5 K for selected compositions in $0 \leqq x \leqq 0.1$. (d)-(i) Magnetoresistance, defined as $\Delta\rho(B)/\rho(0)$ ($\Delta\rho(B) = \rho(B) - \rho(0)$), at 5 K ((d)-(f)) and 100 K ((g)-(i)).

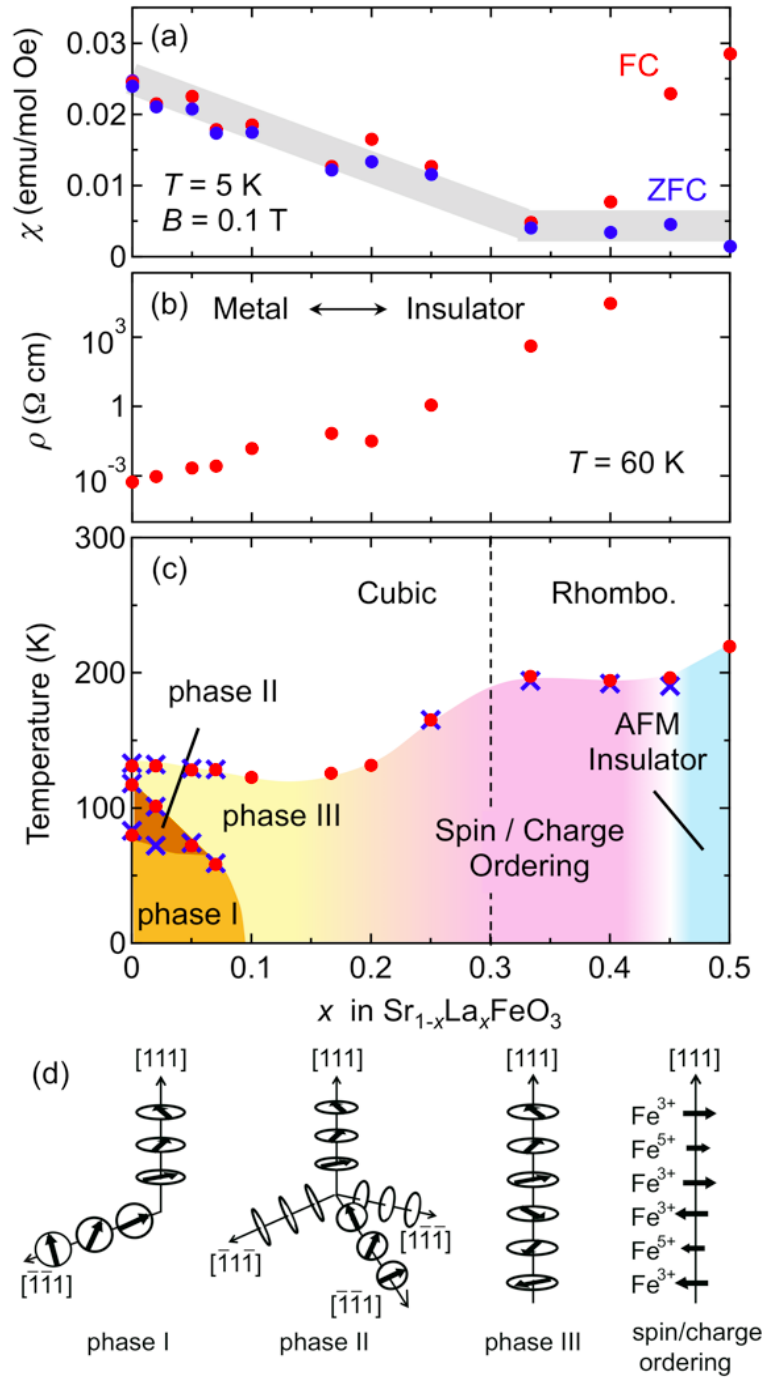

Fig. 6. (a) Composition dependence of magnetic susceptibility at $T = 5$ K measured at $B = 0.1$ T after field cooling (FC) and zero-field cooling (ZFC). (b) Composition dependence of resistivity at $T = 60$ K measured on a warming run at $B = 0$ T. (c) Phase diagram for $Sr_{1-x}La_xFeO_3$. The red circles and blue crosses represent the transition temperatures determined by the warming runs of magnetization and resistivity measurements, respectively. Phases I-III are helimagnetic phases with different spin spirals. (d) Schematic spin image of the magnetic structures for phases I - III and the spin/charge ordered phase. These schematics are drawn assuming the presence of an external magnetic field along the [111] axis.